%% file: main.tex
\newcommand{\name}{SN 2016hil}
\newcommand{\package}[1]{\texttt{#1}}

\documentclass[apj,]{emulateapj}
\usepackage{amsmath}
\usepackage{amssymb}
\usepackage{euscript}
\usepackage{algorithm2e,algorithmic}
\usepackage{bbold}
\usepackage{paralist}
\usepackage{hyperref}
\usepackage{booktabs}

\hypersetup{
    bookmarks=true,         
    unicode=true,          
    colorlinks=true,       
    linkcolor=red,          
    citecolor=blue,        
    filecolor=magenta,      
    urlcolor=red           
}

\def\gtorder{\mathrel{\raise.3ex\hbox{$>$}\mkern-14mu
             \lower0.6ex\hbox{$\sim$}}}
\def\ltorder{\mathrel{\raise.3ex\hbox{$<$}\mkern-14mu
             \lower0.6ex\hbox{$\sim$}}}

\newcommand{\galex}{\textit{GALEX}}

\slugcomment{Draft of \today}

\shortauthors{Irani et al.}

\begin{document}

\title{\name\ -- a Type II supernova in the remote outskirts of an elliptical host and its origin}

\author{Ido Irani\altaffilmark{1}, Steve Schulze\altaffilmark{2}, Avishay Gal-Yam\altaffilmark{1}, Ragnhild Lunnan\altaffilmark{3}, Thomas G. Brink\altaffilmark{4}, WeiKang Zheng\altaffilmark{4}, Alexei V. Filippenko\altaffilmark{4,5}, Yi Yang\altaffilmark{2}, Thomas de Jaeger\altaffilmark{4}, Peter E. Nugent\altaffilmark{4,6}, Mansi M. Kasliwal\altaffilmark{7}, Christoffer Fremling\altaffilmark{7},   James Don Neill\altaffilmark{7}, Umaa Rebbapragada\altaffilmark{8}, Frank J. Masci\altaffilmark{9}, Jesper Sollerman\altaffilmark{3}, Ofer Yaron\altaffilmark{1}
}

\altaffiltext{1}{Department of Particle Physics and Astrophysics, Weizmann Institute of Science, Rehovot 76100, Israel}
\altaffiltext{2}{Benoziyo Center for Astrophysics, The Weizmann Institute of Science, Rehovot 76100, Israel}
\altaffiltext{3}{The Oskar Klein Centre \& Department of Astronomy, Stockholm University, AlbaNova, SE-106 91 Stockholm, Sweden}
\altaffiltext{4}{Department of Astronomy, University of California, Berkeley, CA 94720-3411, USA}
\altaffiltext{5}{Miller Senior Fellow, Miller Institute for Basic Research in Science, University of California, Berkeley, CA 94720, USA}
\altaffiltext{6}{Lawrence Berkeley National Laboratory, Berkeley, California 94720, USA}
\altaffiltext{7}{Division of Physics, Math, and Astronomy, California Institute of Technology, Pasadena, CA 91125, USA}
\altaffiltext{8}{Jet Propulsion Laboratory, California Institute of Technology, Pasadena, CA 91109, USA}
 \altaffiltext{9}{IPAC, California Institute of Technology, 1200 E. California Boulevard, Pasadena, CA 91125, USA}

\begin{abstract}
Type II supernovae (SNe) stem from the core collapse of massive  ($>8\ M_{\odot}$) stars. Owing to their short lifespan, we expect a very low rate of such events in elliptical host galaxies, where the star-formation rate is low, and which mostly consist of an old stellar population. \name\ (iPTF16hil) is a Type II supernova located in the extreme outskirts of an elliptical galaxy at redshift $z=0.0608$ (projected distance $27.2$ kpc). It was detected near peak brightness ($M_{r} \approx -17$ mag) 9 days after the last nondetection. \name\ has some potentially peculiar properties: while presenting a characteristic spectrum, the event was unusually short lived and declined by $\sim 1.5$ mag in $< 40$ days, following an apparently double-peaked light curve. Its spectra suggest a low metallicity ($Z<0.4\ Z_{\odot}$). We place a tentative upper limit on the mass of a potential faint host at $\log(M/M_{\odot}) =7.27^{+0.43}_{-0.24}$  using deep Keck optical imaging. In light of this, we discuss the possibility of the progenitor forming locally, and other more exotic formation scenarios such as a merger  or common-envelope evolution causing a time-delayed explosion. Further observations of the explosion site in the ultraviolet are needed in order to distinguish between the cases. Regardless of the origin of the transient, observing a population of such seemingly hostless Type II SNe could have many uses, including an estimate the number of faint galaxies in a given volume, and tests of the prediction of a time-delayed population of core-collapse SNe in locations otherwise unfavorable for the detection of such events. 

\end{abstract}
\section{Introduction}

The progenitors of Type II supernovae (SNe) are recognized to be massive stars ($>8\ M_{\odot}$; e.g., \citealt{smartt2009}) at the end of their lives. Owing to their short lifespan, we expect a very low rate of such events far from star-forming   regions \citep{Anderson2006}, and in particular in early-type galaxies, which mostly consist of an old stellar population (i.e., of low-mass stars).  Indeed, a systematic analysis of the hosts of SNe \citep{Hakobyan2012} reveals no core-collapse SNe (CCSNe) in elliptical (E) hosts, and only two cases in lenticular (S0) hosts, in comparison to 147 Type Ia SNe in such galaxies from the same sample. The few hosts of Type II/Ib SNe previously thought to be early-type galaxies were misclassified according to this analysis.  Another analysis of these debated cases  \citep{suh2011} demonstrates a systematically bluer color and stronger radio emission of the supposed early-type hosts of CCSNe compared to early-type hosts of SNe Ia -- signatures of recent star formation (SF). More generally, a fraction of early-type galaxies have demonstrated some SF (see, e.g., \citealt{Crocker2011}, \citealt{Kaviraj2007}, or \citealt{Kaviraj2008}). It has been suggested that minor mergers are the main mechanism of such residual star formation \citep{Kaviraj2009}. This gives reasons to expect a residual rate of CCSNe in early-type galaxies.\\

There are also possible reasons to expect a residual population of CCSNe in regions with no recent SF. \cite{zapartas2017} suggest that a significant fraction ($\sim 15\%$) of CCSNe are caused by mass transfer between a pair of intermediate-mass (4--8 $M_{\odot}$) binaries, occurring up to 200 Myr after stellar birth (``late" events). Similarly, \cite{soker2019} outlines several possible mechanisms through which common-envelope evolution may terminate in CCSNe. Such scenarios would involve a secondary star or stripped core spiraling into the envelope of a larger primary star, resulting in a SN explosion. \\   

From an observational point of view, there have been rare cases of non-Ia SNe in early-type hosts, where no nearby star formation could be measured. For example, the Type Ibn SN PS1-12sk \citep{sanders2013} occurred in the local environment of an E host (projected separation $28.1$ kpc). \cite{Hosseinzadeh2019} analyse deep ultraviolet (UV) images of the event obtained with the {\it Hubble Space Telescope (HST)}, and find no measurable SF activity in the region. In light of this, it has been suggested that the progenitor of PS1-12sk might not have been a massive star, and some alternatives have been suggested. 
Similarly, ``Ca-rich" SNe Ib \citep{Filippenko2003} are thought to be the product of interactions between two white dwarfs and not a result of core collapse \citep{Perets2010,Waldman2011}, as they essentially always occur in or near old stellar environments \citep{Lunnan2017}. For example, the environment of SN 2005cz \citep{Kawabata2010} was investigated thoroughly and demonstrated to exhibit no star formation \citep{perets2011}. \\

Finally, early-type galaxies may have dwarf satellites which do present some SF activity. Given the limiting magnitudes ($\sim22.5$ mag) of the Sloan Digital Sky Survey (SDSS; \citealt{york2000}) and the Panoramic Survey Telescope and Rapid Response System 1 (PS1; \citealt{Chambers2016}), we would be unable to detect galaxies fainter than $M \approx -14.5$ mag at a redshift $z=0.06$, which is relevant for this study. We certainly expect a non-negligible fraction of CCSNe to occur in such hosts \citep{Arcavi2010}. 


In the past decade, automated and systematic surveys have increased by orders of magnitude the rate of SN discoveries. These include the Palomar Transient Factory (PTF; \citealt{law2009}), its inheritor the intermediate Palomar Transient Factory (iPTF; \citealt{kulkarni2013}), the All Sky Automated Survey for Supernovae (ASAS-SN; \citealt{shappee2014}), the Asteroid Terrestrial-impact Last Alert System (ATLAS; \citealt{Tonry2018}), PS1, and the {\it Gaia} Photometric Science Alerts \citep{Wyrzykowski2012}. For example, (i)PTF discovered more than $4000$ SNe, of which $\sim950$ were core-collapse events. This provides access to populations which occur at a rate of $\sim 1\%$ of all core-collapse events. Within the next year, the Zwicky Transient Facility (ZTF; \citealt{Bellm2019}) is expected to observe a similar number of events. This will open a window for studying new and exotic populations of transients, of which few events were observed in the past or that are completely unknown.  

Here we present the case of \name. The event was discovered \citep{Kasliwal2018a} and classified \citep{Irani2019a} by iPTF as a spectroscopically regular Type II SN \citep[e.g.,][]{Filippenko1997,galyam2017}. \name\ occurred in an unusual location -- the outskirts of an E galaxy. We describe our spectroscopic and photometric observations in Sect. 2, present  our findings concerning the transient and its host galaxy in Sect. 3, and discuss possible origins for the event in Sect. 4. Throughout this paper we assume $H_0=67.11~{\rm km\,s}^{-1}\,{\rm Mpc}^{-1}$ and a $\Lambda$CDM cosmology with $\Omega_m=0.32$ and $\Omega_\Lambda=0.68$ \citep{Planck2014a}.\\

\begin{figure}[h]
\centering
\includegraphics[width=1\columnwidth]{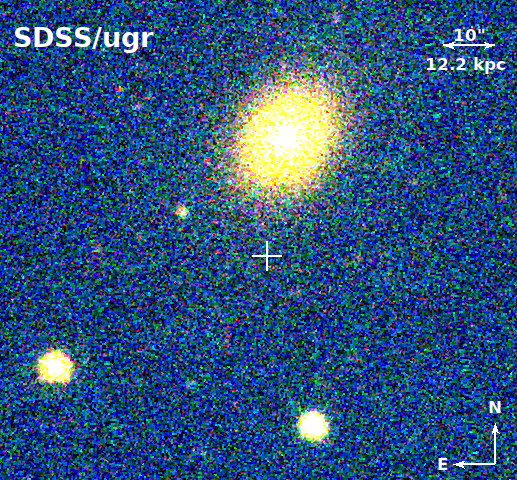}
\caption{The host galaxy of \name\ as observed by SDSS in late 2004 in the \textit{ugr} bands. The event location is marked with a white cross. \name\ was observed $23.1"\pm0.3"$ from the region of maximal brightness in the host, corresponding to a projected separation of $27.2\pm0.4$  kpc, assuming the host redshift of $z=0.06079$.}  
\label{fig:host}
\end{figure}

\begin{figure*}[t]
\centering
\includegraphics[width=2\columnwidth]{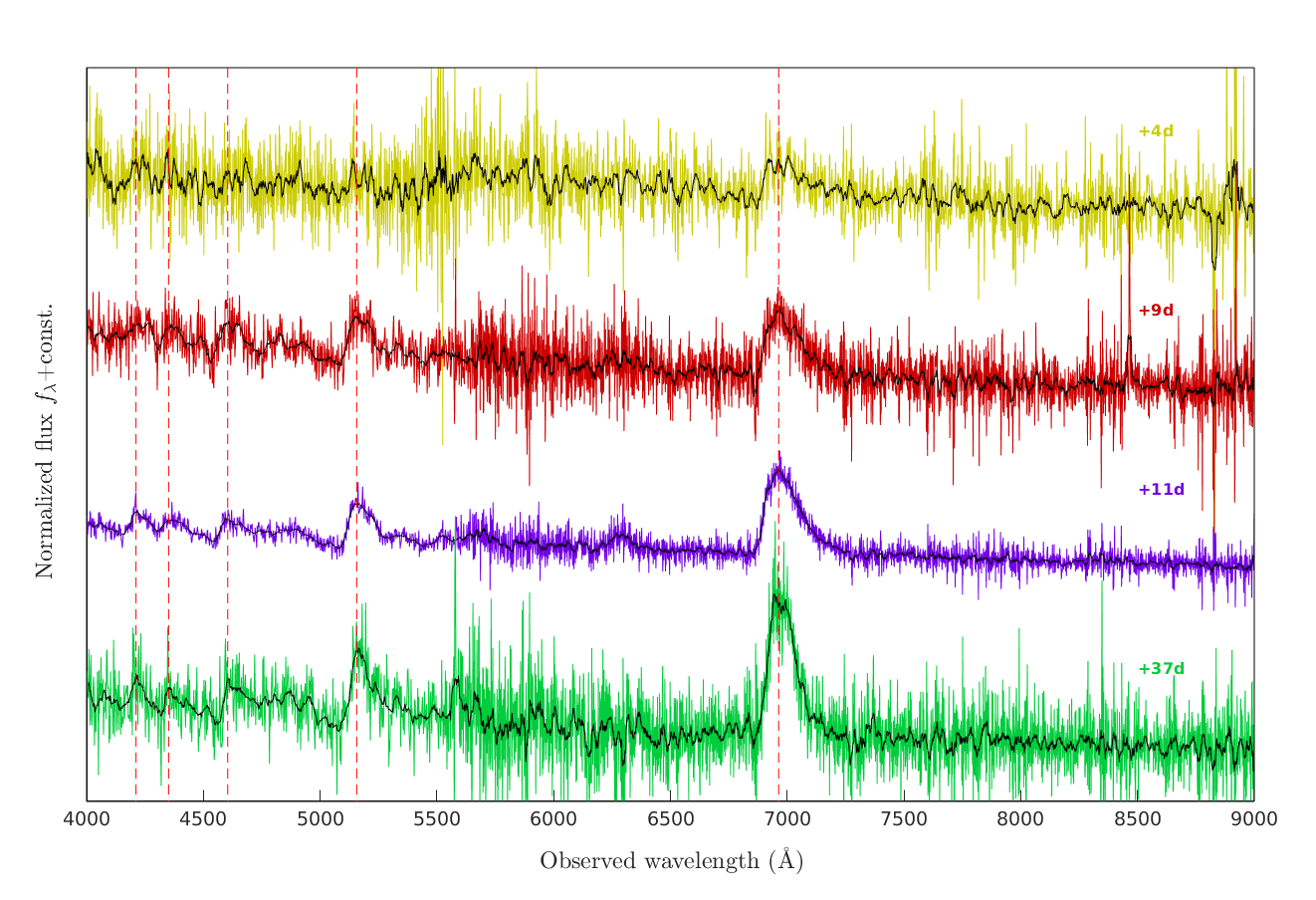}
\caption{Spectral evolution of \name. The spectra (color) are overlaid with a smooth version (black), and labeled according to their observation time relative to first detection. Red dashed lines correspond to redshifted hydrogen lines H$\alpha$\ through H$\varepsilon$ (from right to left). Spectra are trimmed below 4000 \AA\ owing to low signal-to-noise ratio (S/N) at these wavelengths.} 
\label{fig:spectra}
\end{figure*}

\begin{figure*}[t]
\centering
\includegraphics[width=2\columnwidth]{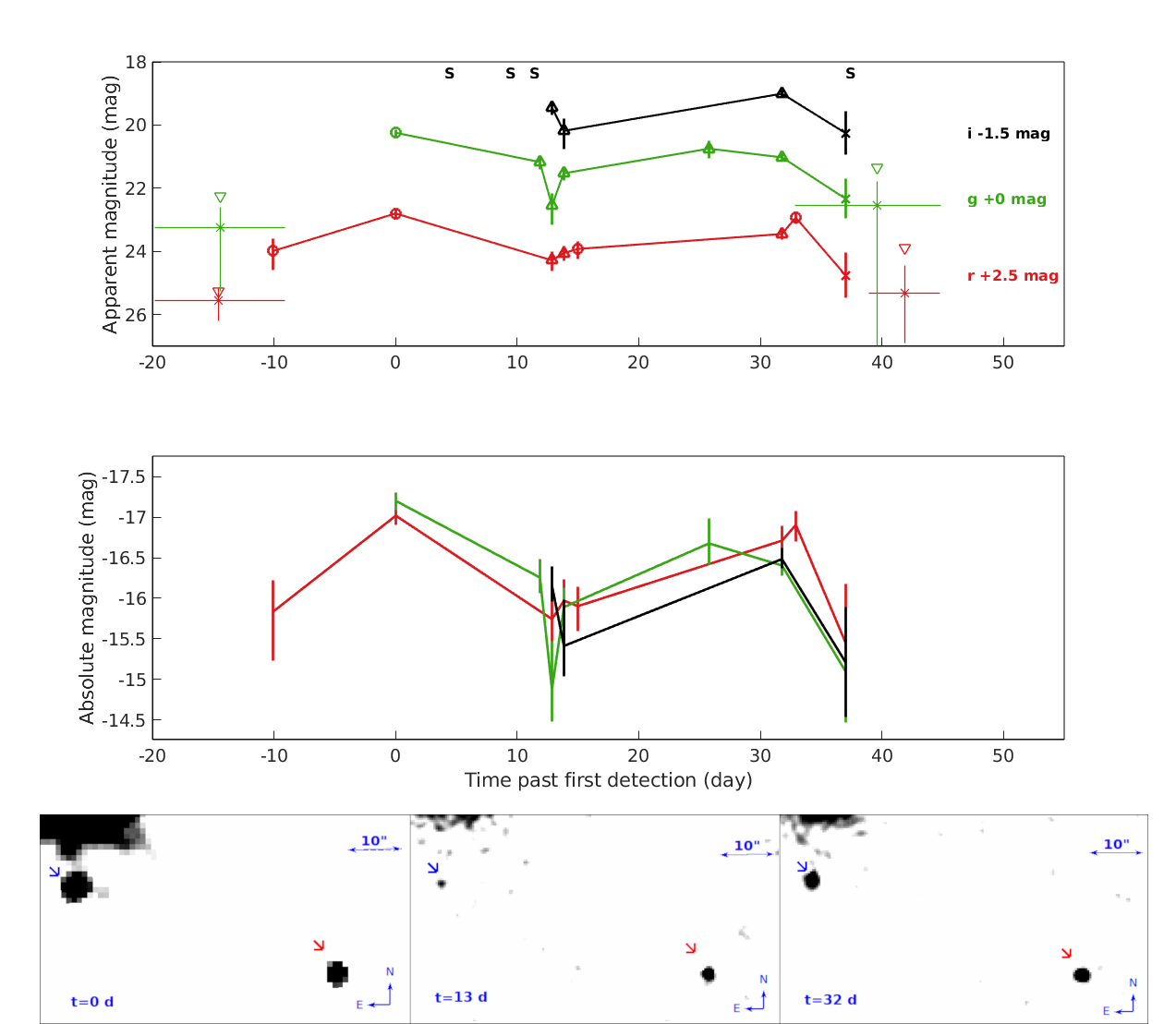}

\caption{In the top panel are optical light curves compiled from P48, P60, and synthetic photometry from the late-time Keck/LRIS spectrum, demonstrating the photometric evolution of \name. The data-point markers correspond to the instrument used: circles are P48, upward-pointing triangles are P60/SEDM, and x are synthetic photometry. The points on the sides of the light curves correspond to summed fluxes (stars) and derived $3\sigma$ limits (downward-pointing triangles). The period over which the data were stacked is indicated by the horizontal error bars. In the middle panel, absolute magnitudes are plotted separately owing to the significant (up to 0.3 mag) difference in S-correction. Smoothed observations in the $r/R$ bands are presented in the lower panel for reference for $t=0$, 13, and 32 d. A red arrow points to a nearby source (which is not variable compared to other stars in the field), and a blue arrow points to the location of \name. Notice the visible decline in brightness at $t=13$ d, followed by an increase in brightness at $t=32$ d.} 
\label{fig:photometry}
\end{figure*}

\section{Observations}
\label{sec:observations}


\subsection{Discovery, classification, and host galaxy}
\name\ was detected using the 48-inch  telescope at Palomar Observatory  (P48), on 2016 October 22 (UT dates are used throughout this paper) at 07:55 (JD 2457683.801), in the \textit{r} band ($20.29\pm0.12$ mag) and was observed in the \textit{g} band ($20.34\pm0.13$ mag) 40 min later. The source was at $\alpha = 01^{\rm h}10^{\rm m}24.75^{\rm s}$, $\delta = +14^\circ12'15.5''$  (J2000). The last nondetection was 9 days before the explosion down to a $3\sigma$ limit of 20.89 mag in the \textit{r} band, although an earlier marginal detection was later identified (see Sect. \ref{sec:photometry}). 

In Fig.  \ref{fig:host}, we present the unusual location of the event: in the outskirts of the E galaxy SDSS J011024.51+141238.7. This galaxy is observed at a redshift $z=0.06079$, consistent with the redshift derived from the H$\alpha$ emission line in the spectra of \name\ (see Sect. \ref{sec:spectroscopy}).

\subsection{Optical Spectroscopy}
\label{sec:spectroscopy}

We collected four optical spectra during a period of 40 days when \name\ was visible. On 2016 October 26, the first spectrum of the SN was obtained using the Double Beam Spectrograph (DBSP; \citealp{Oke1982}) mounted on the Palomar 200-inch Hale telescope (P200). The gratings of 600/4000 and 316/7500 were used for the blue and red cameras, respectively, with the D55 dichroic. The data were reduced using standard procedures, including bias and flatfield corrections, one-dimensional (1D) spectral extraction, wavelength calibration with comparison lamps, and flux calibration using observations of standard stars observed during the same night and at approximately similar airmasses to the SN.

Three additional spectra were obtained with the Low Resolution Imaging Spectrometer (LRIS; \citealt{Oke1995}) on the 10-m Keck I telescope. The gratings of 300/3400 and 300/8500 were used for the blue and red cameras, respectively, with the 560 dichroic. The data were reduced using the  \package{LRIS automated reduction pipeline} (\package{LPipe})  \citep{Perley2019}, and are made available to the public on WISeREP \citep{yaron2012}.  
In Table \ref{tab:opt-spec}  we report our spectral observation log and in Fig.  \ref{fig:spectra} we present the full set of spectra. They exhibit broad hydrogen emission features which evolve rapidly throughout the observation period, based on which \name\ is classified as a Type II SN.  


\begin{deluxetable}{lclcc} 
\tablecaption{Log of Optical Spectra of \name\ }

\tablewidth{0pt} 
\tablehead{ \colhead{Date} & \colhead{$\Delta t$ (d)$^a$} & \colhead{Instrument} & \colhead{Exp. time (s)} & \colhead{Airmass}} 
\tabletypesize{\scriptsize} 
\startdata 
2016 Oct. 26 & 4 & P200/DBSP & 600$\times$2 & 1.10\\ 
2016 Oct. 31 & 9 & Keck/LRIS & 1850 & 1.35\\ 
2016 Nov. 02 & 11 & Keck/LRIS & 1160 & 1.07\\ 
2016 Nov. 28 & 37 & Keck/LRIS & 870 & 1.41\\ 
\enddata 
\label{tab:opt-spec}
\tablenotetext{a}{Relative to first detection.}
\end{deluxetable}

\subsection{Optical Photometry}
\label{sec:photometry}


After the detection, follow-up observations were made using the Spectral Energy Distribution Machine (SEDM; \citealt{Blagorodnova2018a}) mounted at the 60-inch telescope at Palomar Observatory (P60) in addition to routine monitoring with the P48. Photometry was acquired with SDSS \textit{g} and Mould-$R$ bands for the P48 images, and with SDSS \textit{gri} bands for the P60 images. Mould-$R$ was then converted to the SDSS \textit{r}  band using the Lupton color equations (2005).\footnote{\href{https://www.sdss3.org/dr10/algorithms/sdssUBVRITransform.php}{https://www.sdss3.org/dr10/algorithms/sdssUBVRITransform.php}}
Since \name\ is located on a simple background, we chose to use aperture photometry in order to extract source fluxes. This was done by designing custom apertures and annuli  with the MATLAB Astronomy \& Astrophysics Toolbox\footnote{\href{https://github.com/EranOfek/MAAT}{https://github.com/EranOfek/MAAT}} \citep{ofek2014}. We calibrated zeropoints with SDSS stars. Using the images from the days previous to first detection, we summed the nondetection fluxes and derived summed nondetection limits to constrain the shape of the light curve before peak brightness (not including the flux from the marginal detection at $t=-9$ d). We repeated this procedures for the epochs after rebrightening observed at $t\approx 32$ d, when poor weather conditions at Palomar prevented further photometric observations. 

Moreover, we obtained approximate photometry synthesized from the Keck/LRIS spectrum taken at 37 days after first detection. To acquire some estimate for the systematic error involved in synthetic photometry, we compared the scatter of the synthetic photometry acquired from the earlier spectra to the linear interpolation of the light curve in the relevant filter. For the \textit{i}-band filter for which no such data exist, we took the error to be the mean of the uncertainty in the \textit{gr} bands. 

We corrected for Galactic extinction using the NASA/IPAC Extragalactic Database\footnote{\href{https://ned.ipac.caltech.edu/extinction\_calculator}{https://ned.ipac.caltech.edu/extinction\_calculator}} (NED), which cites a value of $A_{V}=1.21$ mag for this line of sight based on \cite{schlafly2011}. 

S-corrections \citep{Stritzinger2002} were estimated for the appropriate filters at the times of the spectra, and by then linearly interpolating the trend for different epochs. This became significant (up to $\sim0.3$ mag) for the P60 \textit{r}-band photometry since at the redshift of \name\ the evolving H$\alpha$ feature is at the boundary of the filter (see Fig. \ref{fig:spectra}). Absolute magnitude light curves are thus plotted separately.

Table \ref{tab:opt-phot} reports the measured the \textit{gri} magnitudes for the Palomar data, as well as the late-time photometry from Keck in Sect. \ref{sec:keck_observations}. The \textit{gri} S-corrected light curve is presented in Fig. \ref{fig:photometry}. Photometry is made available on WISeREP. Although the S/N is low, we tentatively suggest that the light curve of \name\ clearly has a double peak, which can also be corroborated by the lower panel of Fig. \ref{fig:photometry}. \\
\\

\begin{deluxetable}{cllcc} 

\tablecaption{Ground-based optical photometry of \name} 
\tablehead{\colhead{$\Delta t$ (d)$^a$ } & \colhead{Instrument} & \colhead{Filter} & \colhead{AB Mag} & \colhead{BC (mag)$^b$}} 
\tabletypesize{\scriptsize} 
\startdata 
-10.07 & P48                   & \textit{r} & 21.49 $\pm$ 0.61 & 0.055$^c$\\ 
0.00   & P48                   & \textit{r} & 20.30 $\pm$ 0.12 & \\ 
4.00   & P200/DBSP$^{\dagger}$ & \textit{r} & 21.04 $\pm$ 0.73 & \\ 
9.00   & Keck/LRIS$^{\dagger}$ & \textit{r} & 20.17 $\pm$ 0.73 & \\ 
11.00  & Keck/LRIS$^{\dagger}$ & \textit{r} & 20.97 $\pm$ 0.73 & \\ 
12.86  & P60/SEDM                   & \textit{r} & 21.78 $\pm$ 0.34 & \\ 
13.82  & P60/SEDM                   & \textit{r} & 21.55 $\pm$ 0.26 & \\ 
14.99  & P48                   & \textit{r} & 21.42 $\pm$ 0.31 & \\ 
31.79  & P60/SEDM                   & \textit{r} & 20.95 $\pm$ 0.18 & \\ 
32.91  & P48                   & \textit{r} & 20.43 $\pm$ 0.21 & \\ 
37.00  & Keck/LRIS$^{\dagger}$ & \textit{r} & 22.25 $\pm$ 0.73 & \\ 
246.29 & Keck/LRIS & \textit{r} & 25.32 $\pm$ 0.50 & \\ 
771.05 & Keck/LRIS$^{d}$ & \textit{r} & 25.96 $\pm$ 0.83 & \\ 
0.03 & P48 & \textit{g} & 20.24 $\pm$ 0.12 & 0.055\\ 
4.00 & P200/DBSP$^{\dagger}$ & \textit{g} & 21.64 $\pm$ 0.63 & \\ 
9.00 & Keck/LRIS$^{\dagger}$ & \textit{g} & 20.35 $\pm$ 0.63 & \\ 
11.00 & Keck/LRIS$^{\dagger}$ & \textit{g} & 21.26 $\pm$ 0.63 & \\ 
11.86 & P60/SEDM & \textit{g} & 21.17 $\pm$ 0.23 & 0.051\\ 
12.87 & P60/SEDM & \textit{g} & 22.55 $\pm$ 0.59 & 0.051\\ 
13.82 & P60/SEDM & \textit{g} & 21.53 $\pm$ 0.24 & 0.050\\ 
25.78 & P60/SEDM & \textit{g} & 20.75 $\pm$ 0.31 & 0.041\\ 
31.79 & P60/SEDM & \textit{g} & 21.02 $\pm$ 0.13 & 0.035\\ 
37.00 & Keck/LRIS$^{\dagger}$ & \textit{g} & 22.33 $\pm$ 0.63 & 0.028\\ 
246.29 & Keck/LRIS & \textit{g} & 26.44 $\pm$ 0.46 & -0.947\\ 
771.05 & Keck/LRIS$^{d}$ & \textit{g} & 27.15 $\pm$ 0.73 & \\ 
4.00 & P200/DBSP$^{\dagger}$ & \textit{i} & 20.91 $\pm$ 0.68 & \\ 
9.00 & Keck/LRIS$^{\dagger}$ & \textit{i} & 19.91 $\pm$ 0.68 & \\ 
11.00 & Keck/LRIS$^{\dagger}$ & \textit{i} & 20.57 $\pm$ 0.68 & \\ 
12.87 & P60/SEDM & \textit{i} & 20.95 $\pm$ 0.24 & \\ 
13.82 & P60/SEDM & \textit{i} & 21.68 $\pm$ 0.57 & \\ 
31.79 & P60/SEDM & \textit{i} & 20.51 $\pm$ 0.13 & \\ 
37.00 & Keck/LRIS$^{\dagger}$ & \textit{i} & 21.75 $\pm$ 0.68 & \\ 
246.29 & Keck/LRIS & \textit{r}+\textit{g} & 24.99 $\pm$ 0.41 & \\ 
\enddata 
\tablenotetext{$\dagger$}{Synthetic photometry.}
\tablenotetext{a}{Relative to first detection.}
\tablenotetext{b}{Bolometric correction.}
\tablenotetext{c}{Applied on \textit{g}-band photometry derived from color fit. See details in Sect. \ref{sec:analysis}.}
\tablenotetext{d}{2 $\sigma$ measurements used for limits in Sect. \ref{sec:host}}

\label{tab:opt-phot}

\end{deluxetable}

\subsection{Late-time observations}
\label{sec:keck_observations}

\begin{figure}[t]
\centering
\includegraphics[width=1\columnwidth]{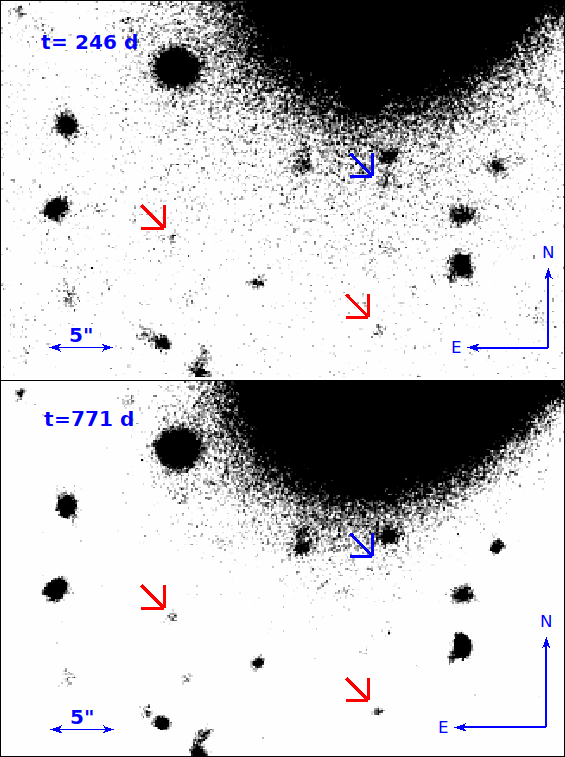}
\caption{Deep Keck/LRIS \textit{r}+\textit{g} observations of the event location at 246 (top panel) and 771 (bottom panel) days after detection (JD 2458454 and JD 2457930, respectively). In both panels, blue arrows  point at the location of the event, and red arrows point at sources of comparable brightness for reference. Even though the detection is marginal in the \textit{r} and \textit{g} bands separately, it becomes significant when viewed in the \textit{r}+\textit{g} summed image.} 
\label{fig:keck}
\end{figure}

On the nights of 2017 June 24 ($t=246$ d) and 2018 December 1 ($t=771$ d) we obtained simultaneous \textit{r} and \textit{g} photometry of \name\ with Keck/LRIS.  The 2017 June data consist of four dithered exposures totaling 1290 s in \textit{g} and 1200 s in \textit{r}.  The 2018 December data consist of eight dithered exposures totaling 2598 s in \textit{g} and 2400 s in \textit{r}.  These data were processed following standard techniques for CCD reductions using \package{LPipe}.

In order to eliminate contamination by residual light from the nearby galaxy and from surrounding sources, aperture photometry was performed manually: background and background noise were estimated by establishing an elliptical contour of the host and extending it to reach the location of the event. A series of custom apertures (with a radius of $1.27''$) were then constructed along this contour, and the background flux was measured with adjustments for any additional flux gradient. The manual measurements were performed in \package{SAOImageDS9} \citep{Joye2003}. 
The photometric zeropoints were acquired using unsaturated stars in the field and by comparing them to the converted SDSS catalog filters as discussed in Sect. \ref{sec:photometry}.
Extinction was treated as discussed in Sect. \ref{sec:photometry}, and no S-corrections were applied.

In the first epoch, there were faint and marginally significant detections of the transient in \textit{r} and \textit{g} separately. In order to boost the significance of the detection, \textit{r} and \textit{g} images were summed, after manual cross-astrometry was performed using the \package{Graphical Astronomy and Image Analysis Tool} (\package{GAIA}; \citealt{Draper2014}). This resulted in a $>3\sigma$   detection in the summed image. 
In Fig. \ref{fig:keck} a comparison between both epochs in the synthetic $R$+\textit{g} band is made, demonstrating the presence of a transient in the first epoch and its absence in the later epoch.\\
\\

\section{Results}
\label{sec:analysis}

\subsection{Light curves}
\name\ has peculiar photometric properties for a spectroscopically regular SN II. These usually present a plateau light curve (IIP) or a linearly declining light curve (IIL); in the most rapid cases, the latter decline by $\sim1.5$\,mag over a period $\geq60$ d \citep{arcavi2017}. The light curve of \name\ is thus unusually short lived, declining by $\sim 1.5$ mag in $<40$ days. The event also presents a double peak in the $gri$ bands, as can be corroborated from the lower panel of Fig. \ref{fig:photometry}. Although not consistent with a plateau or a linear decline, the photometry is quite noisy. It remains to be seen whether these peculiarities will repeat in similar events in the future. For the rest of the paper, we assume the double peak of the light curve is real. However, none of our main conclusions change if this is not the case.

Unusual for a spectroscopically normal SN II, a double-peaked light curve is more characteristic of Type IIb events (see, e.g., \citealt{arcavi2017} for discussion). The spectroscopic features of \name, however, exclude the SN IIb classification since there are no strong helium signatures and prominent presence of hydrogen persists throughout the spectral evolution. In SNe IIb, double-peaked light curves have been suggested to be the result of a peculiar density structure \cite{Bersten2012}. \cite{Nakar2014} show that such a light curve can be produced by a compact core surrounded by an extended low-mass envelope. In some cases the double-peaked structure is attributed to binary interaction (as shown, for example, by \citealt{Benvenuto2013}). On the other hand, \citealp{sapir2016} claim that such a density structure is not necessary to produce a double peak, which can be produced by a standard progenitor star. In such double-peaked light curves, the first peak is generally thought to be powered by shock cooling, and the second peak by the radioactive decay of  $\rm ^{56}Ni$. The double-peaked light curve of \name\ seems to indicate that the event had at least an unusual progenitor. 
 
 \begin{figure}[t]
\centering
\includegraphics[width=1\columnwidth]{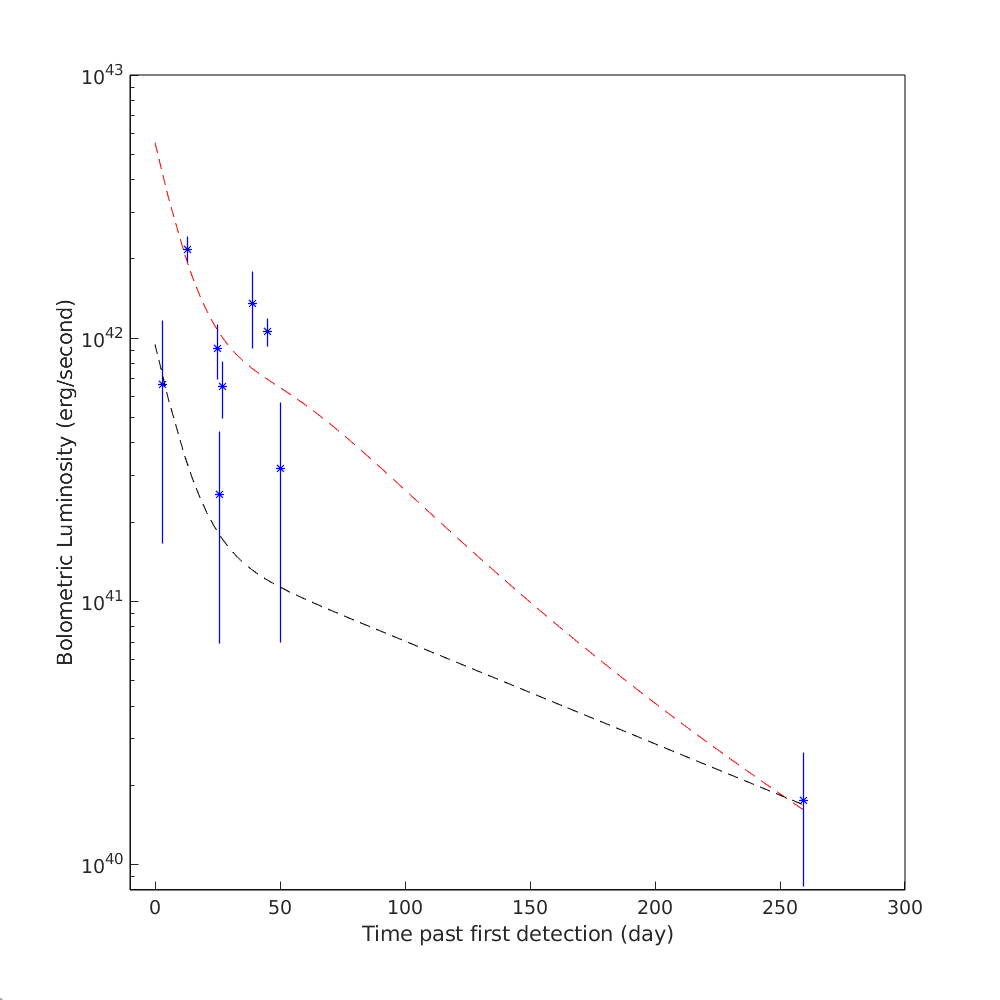}
\caption{Bolometric light curve of \name\ (blue points), plotted together with radioactive-decay output of $^{56}$Ni for two limiting cases: $M_{\rm ^{56}Ni}= 0.07\,M_{\odot}$, $t_{0}=100\,d$ (red dashed line), and $M_{\rm ^{56}Ni}= 0.012\, M_{\odot}$, $t_{0}\rightarrow\infty$ (black dashed line). } 
\label{fig:bollometric}
\end{figure}

\subsection{Bolometric light curve}
We estimated the bolometric light curve of \name\ to see if it is consistent with a radioactively powered light curve, and acquire limits on the corresponding $\rm ^{56}Ni$ mass. 
The bolometric correction was estimated from  \textit{g} magnitudes and the \textit{g-r} color, by using a quadratic fit to the color based on a sample of SNe II as described by \cite{lyman2014}. Since the color evolution was observed to be linear over the entire period of observations, but color was not available for all epochs,  we fit a linear trend and used this fit to compute the bolometric correction for all times where a measurement in either \textit{g} or \textit{r} was available [including the late-time ($t=37$ d) synthetic photometry point]. The bolometric correction as calculated appears in Table \ref{tab:opt-phot}.  Using the bolometric luminosity, the integrated bolometric energy output of the SN is estimated to be $(7.9\pm3.5)\times10^{48}$ erg. 

We assume the following model for $\rm ^{56}Ni\rightarrow^{56}Fe$ decay \citep[see][]{nakar2016, wygoda2019,katz2013}. At early times, all $\gamma$-rays produced in the decay are scattered and deposit their energy in the ejecta. At late times, only a fraction $f_{\gamma}\approx t_{0}^2/t^2$ of the $\gamma$-rays deposit their energy in the ejecta, where $t_{0}$ is the $\gamma$-ray escape time.  A common interpolation for the intermediate times is $f_{\gamma}\approx (1-e^{-t_{0}^2/t^2})$, which captures the correct limits at late and early times. Using this, the total energy output produced by $\rm^{56}Ni$ decay is given by 
\[Q_{\rm Ni}(t)=\frac{M_{\rm ^{56}Ni}}{M_{\odot}}f_{\rm dep}\cdot(6.45 \ e^{-\frac{t}{8.8\ \rm d}}+1.44 \ e^{ -\frac{t}{111.3\ \rm d}})\times 10^{43}\frac{{\rm erg}}{{\rm s}},\]
where $f_{\rm dep}=(0.97f_{\gamma}+0.03)$ is the total fraction of deposited energy due to the radioactive decay, including the energy deposited by positrons. Using this expression, we can place a lower bound on the total $\rm ^{56}Ni$ mass at late times by assuming that all the luminosity at 246 d is due to $\rm ^{56}Co\rightarrow^{56}Fe$ decay, and that $f_{\gamma}=1$. This gives a lower bound of $M_{^{56}Ni}\geq 0.012 M_{\odot} $. Alternatively, we can compute the $M_{\rm ^{56}Ni}$ for a given $t_{0}$.  

We can further note that since $\int\limits _{0}^{t}Q_{\rm Ni}t'dt'\leq\int\limits _{0}^{t}L_{\rm bol}t'dt'$ for all times ($\int\limits _{0}^{t}L(t')t'dt'$ is a conserved quantity, accounting for adiabatic losses), we can place an upper bound on the $^{56}$Ni mass for a given $t_{0}$, using the $^{56}$Ni mass required to power the entire light curve:
\[\frac{M_{\rm ^{56}Ni}}{M_{\odot}}\leq\frac{\int\limits _{0}^{t}L_{\rm bol}t'dt'}{\int\limits _{0}^{t}f_{\rm dep}\cdot(6.45\ e^{-\frac{t'}{8.8d}}+1.44\ e^{-\frac{t'}{111.3d}})\cdot10^{43}\frac{\rm erg}{\rm s}t'dt'}.\]
This gives an upper limit of  $M_{\rm ^{56}Ni}\leq 0.07\, M_{\odot} $, above which the $\rm ^{56}Ni$ mass as measured from late times will not agree with the integrated luminosity. We note that this upper limit is somewhat dependent on the starting point of the integration, but will not change our results or conclusions significantly. For example, changing the explosion time to 5 days earlier than the first photometry point would increase the upper limit by $30\%$ to $0.09\, M_{\odot}$, which is still well within the typical range for SNe II.\\

In Fig. \ref{fig:bollometric} we present the bolometric light curve plotted together with the two limiting cases $M_{\rm ^{56}Ni}= 0.07\, M_{\odot} $, $t_{0}=100$~d   and $M_{\rm ^{56}Ni}= 0.012\, M_{\odot}$, $t_{0}\rightarrow\infty$  for the energy production from $\rm ^{56}Ni\rightarrow^{56}Fe$  decay. We can thus conclude that the late-time photometry of \name\ can provide a $\rm ^{56}Ni$  content consistent with the second peak of the light curve being powered by   $\rm ^{56}Ni$  decay.

\begin{figure*}[t]
\centering
\includegraphics[width=2\columnwidth]{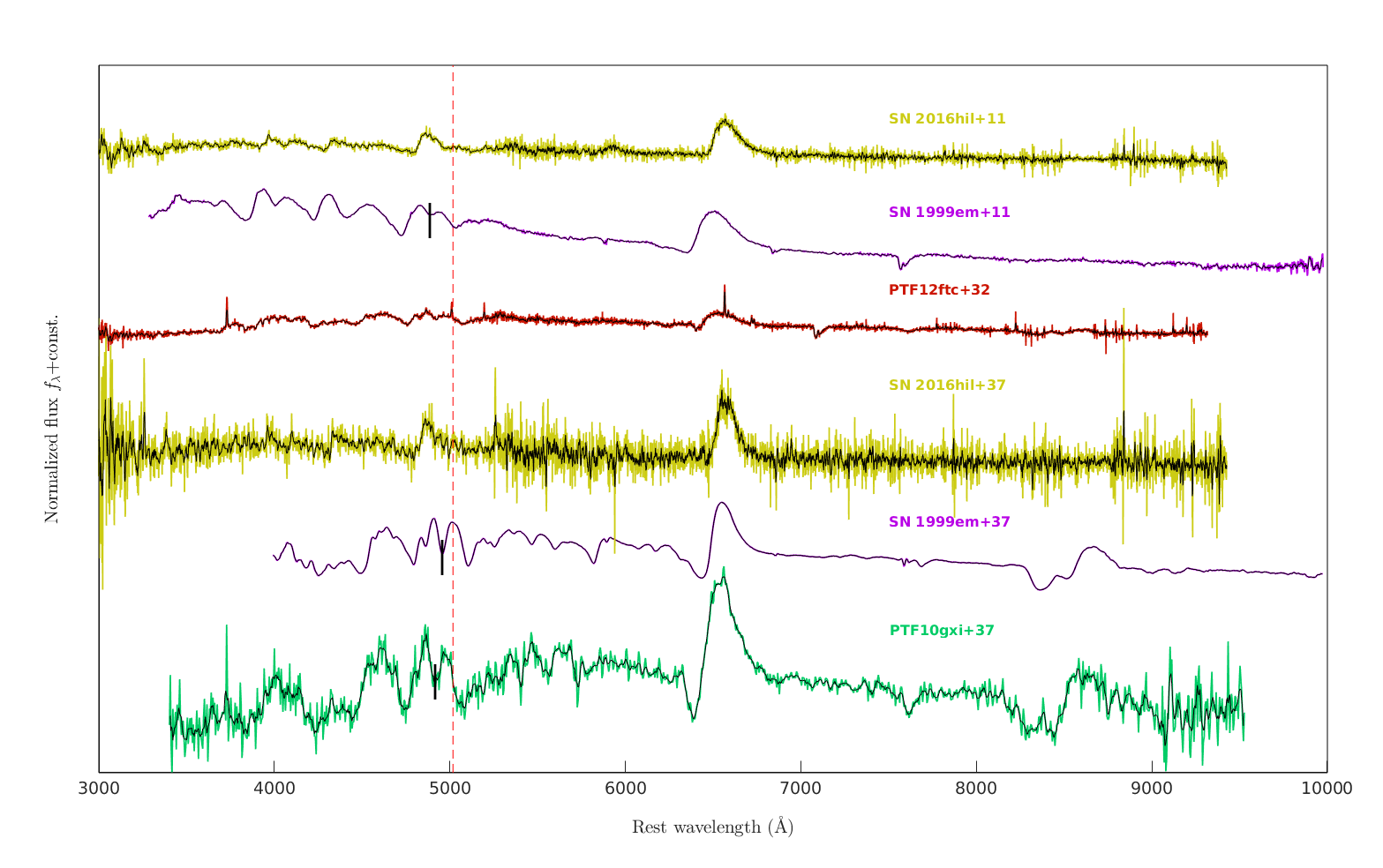}
\caption{Spectra of \name\ at 11 d and 37 d after peak brightness compared with other SNe II at similar phases. Each spectrum is plotted together with a smoothed counterpart (solid black curves).  The dashed red line is the Fe II $\lambda 5018$ line at rest wavelength. The absorption minimum is marked with a solid black line in spectra where the feature is visible.} 
\label{fig:1999em}
\end{figure*}

\subsection{Spectral properties}
As can be seen in Fig. \ref{fig:spectra}, spectra of \name\ display a strong presence of hydrogen, but few other features were identified. \name\ can thus be classified as a spectroscopically regular SN II.  The absorption minima of the P-Cygni profile of the $H\alpha$ feature correspond to expansion velocities of $\sim 5000$ km s$^-1$  throughout the spectral evolution. Across all spectra,  this $H\alpha$  absorption minimum is weak relative to those of other Blamer features. This is more characteristic of a SN IIL than of a SN IIP  (see, e.g., Fig. 18 in \citealt{arcavi2017}). In all spectra, there are no indications of narrow host emission lines, which could serve as indicators of SF. In the spectrum taken 11 d after detection, an unidentified broad emission feature appears near 6300 \AA. It is not seen in the spectrum taken 2 d earlier, probably owing to the low S/N. The lack of other features seems to indicate a low metallicity, which is expected from a low-luminosity host galaxy (i.e., according to the mass-$Z$ relation; \citealt{Tremonti2004}). However, since a metallicity gradient is present in many galaxies (e.g., \citealt{sanchez2014}) the low metallicity of the event could also be consistent with the environment in the outskirts of the main host galaxy.

In a sample by \cite{taddia2016}, the strength of the Fe II $\lambda 5018$ feature was used to determine the metallicity according to the method of \cite{dessart2014}. In Fig. \ref{fig:1999em} we put the spectra of \name\ in context of such SNe, including PTF10gxi and PTF12ftc for which the metallicity was determined to be $Z=0.4\ Z_{\odot}$. The fact that the Fe II $\lambda 5018$ feature  is visible in the spectra of both SNe, and not in any of the spectra of \name, suggests that it has a similar or lower metallicity content \citep[e.g.,][]{Anderson2016,Anderson2018}.

\begin{deluxetable}{lcc} 
\tablecaption{Blackbody fits for  \name\ }
\tablewidth{0pt} 
\tablehead{ \colhead{Date} & \colhead{$\Delta t$ (d)} & \colhead{Temperature (K)}} 
\tabletypesize{\scriptsize} 
\startdata 
2016 Oct. 26 & 4 & 6462 $\pm$  40 \\ 
2016 Oct. 31 & 9 & 7648 $\pm$ 38 \\ 
2016 Nov. 02 & 11 &  6709 $\pm$ 16 \\ 
2016 Nov. 28 & 37 & 7134  $\pm$ 46 \\ 
\enddata 
\tablenotetext{a}{Relative to first detection.}

\label{tab:blackbody}
\end{deluxetable}

The continua of the spectra were fitted to blackbody emission. This was done by iteratively fitting a continuum, subtracting it, removing outliers, and refitting the remaining data, until the temperature converges. In all spectra, the temperature was found to be close to 7000 K, without a clear trend in time. Uncertainties were estimated using 68\% confidence bounds, not accounting for systematic errors. The fitted temperatures and their corresponding uncertainties appear in Table  \ref{tab:blackbody}.

\subsection{Host galaxy}
\label{sec:host}

Identifying the host of \name\ with certainty is crucial for putting this event in context. Our initial association of \name\ with the galaxy SDSS J011024.51+141238.7 is primarily due to \name\ having a redshift consistent with that of the nearby galaxy. We compared the host spectrum, acquired from the SDSS Science Archive Server (SAS), to templates  of various galaxy types \citep{Kinney1996}. It is most consistent with being an E galaxy.

To put this host in context of the general population of host galaxies of SNe II, we compare its photometric properties to the host galaxies of the (i)PTF CCSN sample (Schulze et al., in prep.). This homogeneous sample consists of over 520 SNe II, detected between the beginning of 2009 and the beginning of 2017. We retrieved archival images of the host galaxy from \textit{Galaxy Evolution Explorer} (\galex) Data Release (DR) 8/9 \citep{Martin2005a}, SDSS DR9 \citep{Ahn2012a}, PS1 DR1 \citep{Chambers2016}, the Two-Micron All Sky Survey \citep[2MASS;][]{Skrutskie2006a}, and the unWISE \citep{Lang2014a} images from the NEOWISE \citep{Meisner2017a} Reactivation Year 3. Furthermore, we use the matched-aperture photometry software package \package{Lambda Adaptive Multi-Band Deblending Algorithm in R} \citep[\package{LAMBDAR};][]{Wright2016a} that is based on a photometry software package developed by \citet{Bourne2012a} and tools which will be presented by Schulze et al. (in prep.). The photometry was either calibrated against zeropoints (\galex, PS1, SDSS, and NeoWISE) or against a set of stars (2MASS). The resulting photometry is summarized in Table \ref{tab:host_phot}.

As for the (i)PTF CCSN host-galaxy sample, we model the spectral energy distribution (SED) of the host with the software package \package{LePhare}\footnote{\href{http://www.cfht.hawaii.edu/~arnouts/LEPHARE/lephare.html}{http://www.cfht.hawaii.edu/\~{}arnouts/LEPHARE/lephare.html}} version 2.2 \citep{Arnouts1999a,Ilbert2006a} and standard assumptions (\citealt{Bruzual2003a} stellar population-synthesis models with the Chabrier initial mass function \citealt{Chabrier2003a}, an exponentially declining star-formation history and the \citealt{Calzetti2000a} attenuation curve). 

\begin{figure}[t]
\centering
\includegraphics[width=1\columnwidth]{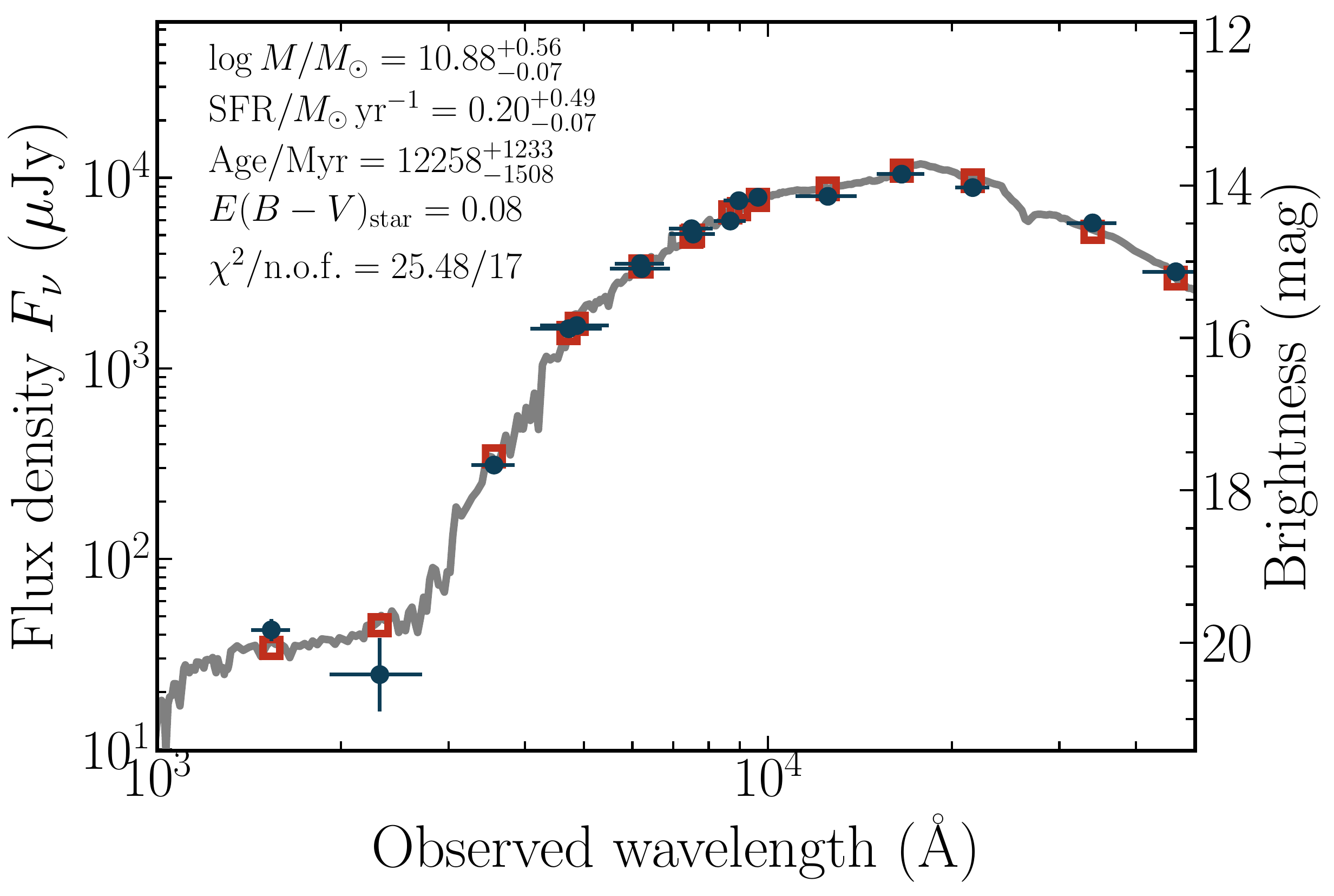}
\caption{Spectral energy distribution of the E galaxy SDSS J011024.51+141238.7 (blue data points) that could have hosted \name. The solid line displays the best fit and the red squares the model-predicted photometry. Key properties of the fit are shown in the figure. The quality of the fit is expressed by the $\chi^2$ divided by the number of filters (n.o.f.).}  
\label{fig:sed}
\end{figure}

\input{tab_host_alex.tex}

Figure \ref{fig:sed} shows the observed SED. It is best described by a galaxy, dominated by an old stellar population, with a large stellar mass content of $\log(M/M_{\odot})=10.88^{+0.56}_{-0.07}$ and a low SF rate (SFR) of ${\rm SFR}= 0.20^{+0.49}_{-0.07}\ M_{\odot}\ \rm yr^{-1}$. The age of the stellar population and the large mass corroborate the conclusion from the SDSS spectrum that this is indeed an E galaxy. The low but non-negligible SFR is not in conflict with this interpretation. \citet{Schawinski2007a} showed that $\sim30\%$ of a volume-limited sample of luminous E galaxies exhibited signs of recent SF.


Another option could be that \name\ occurred in a faint satellite of the main host, where there is still SF activity. As can be seen in Fig.  \ref{fig:keck}, the relatively deep Keck/LRIS images reveal no obvious dwarf galaxy or star-forming region at the location of \name. Using the low-S/N ($2\sigma$) flux detected in the  $t=771$~d epoch in the \textit{r} and \textit{g} bands, we attempt to constrain the galaxy mass and SFR of a possible dwarf satellite host. We repeated the SED fitting process using the \textit{r} and \textit{g} photometry. The results constrain the presence of a potential dwarf host such that $\log(M/M_{\odot})=7.27^{+0.43}_{-0.24}$, and SFR $\leq 0.01\ M_{\odot}\ \rm yr^{-1}$. As the SED is based only on \textit{g} and \textit{r} photometry, the mass estimate should be regarded as an upper limit.

\begin{figure}[t]
\centering
\includegraphics[width=1\columnwidth]{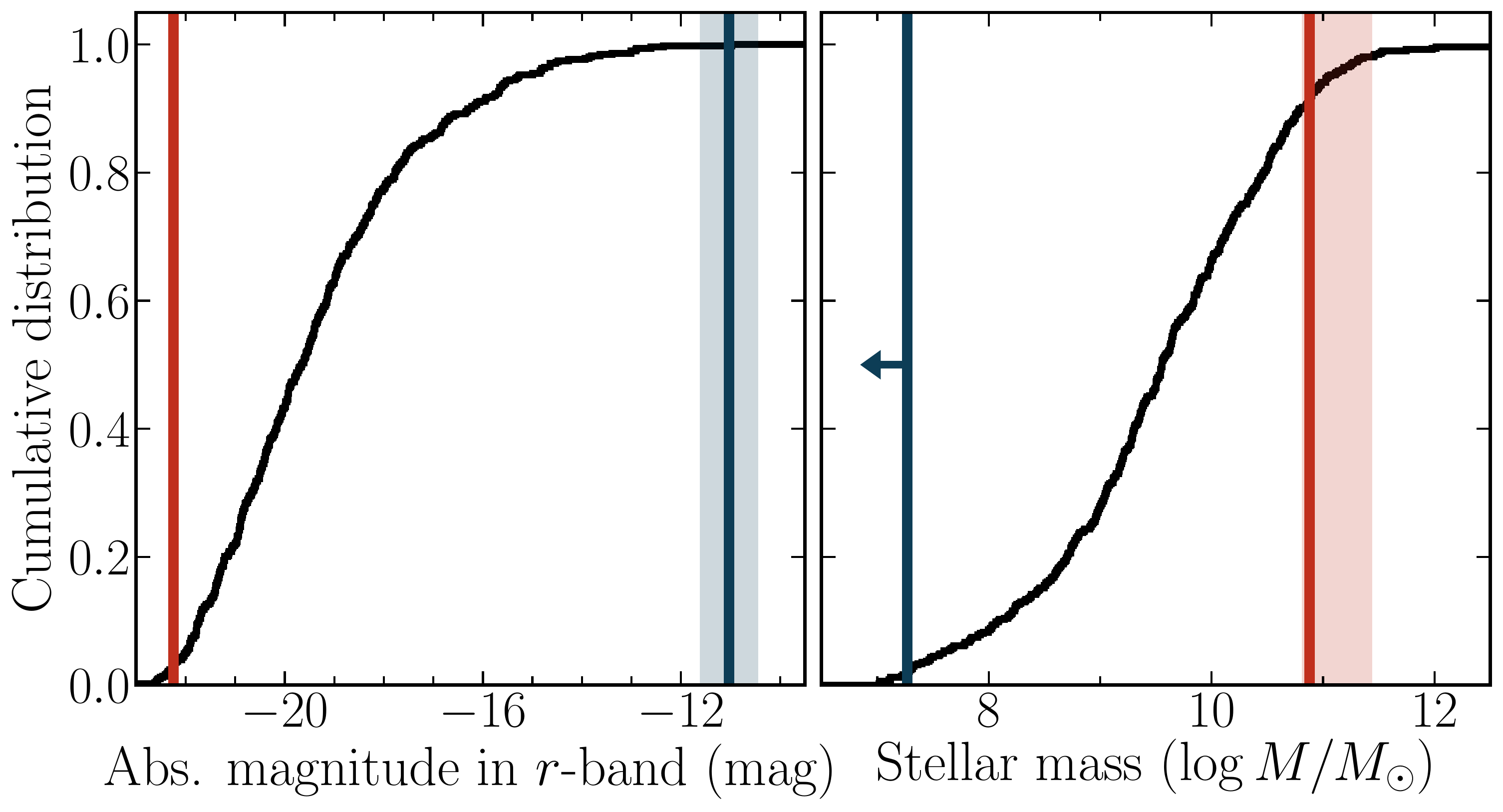}
\caption{Comparison of the host properties (mass and $\textit{r}$-band luminosity) of the two host-galaxy candidates to the general population of SN II host galaxies from the (i)PTF CCSN sample (over 520 objects; black curves). The properties of the E galaxy SDSS J0110+1412 are shown in red and those of the potential dwarf galaxy in blue. The shaded regions indicate the $1\sigma$ uncertainties. Modelling the spectral energy of the dwarf galaxy only provides an upper limit on the stellar mass. This is indicated by the arrow pointing toward lower masses.} 
\label{fig:host_comparison}
\end{figure}

To put both host-galaxy candidates in the context of the general population of SN II host galaxies, we compare their masses and absolute magnitudes to those of the SN II hosts from the (i)PTF survey (Fig. \ref{fig:host_comparison}; values taken from Schulze et al., in prep.). Both candidate host galaxies have extreme values for a SN II host. The E galaxy is among the most luminous and the most massive galaxies in the sample. At the other extreme, the potential dwarf galaxy cospatial with the SN site would be the least luminous host in the SN II (i)PTF sample. Moreover, the mass limit of $10^{7.3}\,M_\odot$ puts this object at the low end of mass functions of star-forming galaxies. 

\section{Discussion and conclusions}
\label{sec:discussion}
Although it is a SN II, \name\ was detected in an unlikely location -- the extreme outskirts of an early-type galaxy, where no SF is expected. The SN presented a low-metallicity spectrum with moderate expansion velocities and blackbody temperatures. Its photometry reveals a double-peaked and short-lived light curve, and late-time photometric observations are consistent with a $\rm ^{56}Ni$ mass sufficient to power the second peak. This being said, the quality of the photometric data is relatively low. It remains to be seen whether similar future events will exhibit comparable properties. 

Deep optical photometry of the environment of \name\ shows no significant sources which could have provided an alternative host where normal SF activity would still be taking place. This, as well as the fact that the nearby galaxy shares the same redshift as \name, make it the immediate candidate for being the host of \name. Still, we do not have enough data to fully exclude the possibility of a very faint host gravitationally bound to the nearby galaxy. That being said, such a dwarf host would have $M_{r}>-12$  mag and an extremely low stellar content of $<10^{7.3}\ M_\odot$. Whether such an extreme host exists could be probed with very deep observations in the visible (below our current Keck limits), and any SF can be best probed by deep observations in the UV from {\it HST}. 

\name\ is thus either a peculiar SN in a normal galaxy, or a peculiar SN in a peculiar galaxy. In either case, this unusual host environment could have interesting implications. We outline several possibilities for the origin of \name. \\

\noindent 
(1) The progenitor was formed in the main part of the nearby E host, and was ejected with high velocity. In such a case, the event is tracing a residual population of massive stars in early-type galaxies -- the result of the remaining high-mass ($>8\ M_{\odot}$) star formation in the host. \cite{Kasliwal2012} show that the vast majority of SNe occur within $10\ \rm kpc$ of the host-galaxy nucleus. A simple calculation demonstrates that such a star would have to travel with a velocity of $\sim 1000\ \rm km\ s^{-1}$ in order to travel $\sim10\ \rm kpc$ within the  $\sim10\ \rm Myr$ of its life span. We tentatively conclude that this option does not seem very likely, as it requires two rare phenomena to occur: residual SF in an E galaxy and a hypervelocity ejection.\\ 

\noindent 
(2) The star was formed locally in a  star-forming satellite of the E galaxy, which still produces high-mass stars. This option is favored as it does not require any modification of the standard paradigm of SN II formation, but is disfavored by the fact that we have strong constraints on the mass and luminosity of a possible host at the location of the event, which can be further tightened in the near future. If this turns out to be the case, \name\ would be the SN II with the faintest host observed by (i)PTF to date. Collecting the statistics of such seemingly hostless SNe could provide a handle on the number of almost invisible faint dwarf galaxies in a given volume and redshift.\\

\noindent 
(3) The progenitor is part of a middle-aged diffuse population of $<8\,M_{\odot}$ stars extending around the host. How can such stars explode as SNe II? Several ideas involving interactions of lower-mass progenitors have been proposed. \cite{zapartas2017} outline evolutionary channels through which ``late" CCSNe (up to 200 Myr after star formation) may occur. One option is that a pair of  main-sequence (MS) intermediate-mass stars (4--8 $M_{\odot}$), or an intermediate-mass MS star and a post-MS star, could merge completely. Such a merger would revive the merger product, which will recover its equilibrium structure and eventually terminate in a CCSN. Other options include the reverse merger of a compact object and a post-MS star, resulting in a CCSN after an initial common-envelope phase, as discussed by \cite{Sabach2014}. 

These binary interaction scenarios could provide a reasonable explanation for a double-peaked light curve -- in the aftermath of a merger, we expect a significant increase in the size of the surrounding envelope. Such an expansion could create a low-mass and extended envelope that could could produce the two peaks. This is reminiscent of our current understanding of SNe IIb, where the envelope of a star is thought to be mostly stripped owing to binary interaction, thus revealing the helium core during its spectral evolution. In this case, however, a hydrogen envelope could remain around the merged core, so that the spectral evolution would remain dominated by hydrogen.\\ 

With the increasing number of SNe detected in the era of automated wide-area transient surveys, new populations of transients are being revealed. We expect that events similar to \name\ will be discovered in the near future, and a population could be established. \name\ shows some potentially peculiar spectroscopic and photometric properties, in addition to its unusual location. Once we discover more \name-like events, we can identify their observational characteristics. These will presumably allow us to answer the question of their origin.

\section{Acknowledgements }
We thank A. Ho and K. De for assistance with some of the observations. A.G.-Y. is supported by the EU via ERC grant No. 725161, the ISF, the BSF Transformative program, and a Kimmel award. A.V.F.'s supernova group at U.C. Berkeley is supported by the TABASGO Foundation, the Christopher R. Redlich Fund, Gary and Cynthia Bengier, and the Miller Institute for Basic Research in Science.

This research has made use of the NASA/IPAC Extragalactic Database (NED),
which is operated by the Jet Propulsion Laboratory, California Institute of Technology,
under contract with the National Aeronautics and Space Administration (NASA). Part of this research was carried out at the Jet Propulsion Laboratory, California Institute of Technology, under a contract with NASA.
Some of the data presented herein were obtained at the W.M. Keck Observatory, which is operated
as a scientific partnership among the California Institute of Technology, the University of California, and NASA; the
Observatory was made possible by the generous financial support of the W.M. Keck Foundation. The authors
wish to recognize and acknowledge the very significant
cultural role and reverence that the summit of Maunakea
has always had within the indigenous Hawaiian community. We are most fortunate to have the opportunity
to conduct observations from this mountain.
Based in part on observations obtained with the 48-inch Samuel Oschin Telescope
and the 60-inch Telescope at the Palomar Observatory as part of the intermediate Palomar Transient Factory (iPTF) project, a scientific collaboration among
the California Institute of Technology, Los Alamos National Laboratory, the
University of Wisconsin, Milwaukee, the Oskar Klein Center, the Weizmann
Institute of Science, the TANGO Program of the University System of Taiwan,
and the Kavli Institute for the Physics and Mathematics of the Universe.

\bibliographystyle{apj} 
\bibliography{main.bbl}

\end{document}

%% file: tab_host_alex.tex
\begin{table}
\centering
\caption{Multiwavelength magnitudes of the host galaxy}\label{tab:host_phot}
\begin{tabular}{lcclcc}
\toprule
Instrument/	    & $\lambda_{\rm eff}$   & Magnitude  	\\
Filter          & (\AA)                 &            \\
\midrule
\galex/FUV		  & 1542     & $ 20.14	\pm 0.11$ \\
\galex/NUV		  & 2274     & $ 20.77	\pm 0.47$ \\
SDSS/$u$ 		  & 3595     & $ 17.85	\pm 0.08$ \\
SDSS/$g$		  & 4640     & $ 16.03	\pm 0.03$ \\
SDSS/$r$ 		  & 6122     & $ 15.13	\pm 0.03$ \\
SDSS/$i$		  & 7440     & $ 14.65	\pm 0.02$ \\
SDSS/$z$		  & 8897     & $ 14.26	\pm 0.03$ \\
PS1/$g_{\rm PS1}$ & 4776     & $ 15.98	\pm 0.03$ \\
PS1/$r_{\rm PS1}$ & 6130     & $ 15.20	\pm 0.01$ \\
PS1/$i_{\rm PS1}$ & 7485     & $ 14.72    \pm 0.01$ \\
PS1/$z_{\rm PS1}$ & 8658     & $ 14.53    \pm 0.02$ \\
PS1/$y_{\rm PS1}$ & 9603     & $ 14.21    \pm 0.02$ \\
2MASS/$J$ 	   	  & 12,482    & $ 14.18    \pm 0.05$ \\
2MASS/$H$ 	   	  & 16,620    & $ 13.87    \pm 0.05$ \\
2MASS/$K_s$    	  & 21,590    & $ 14.04    \pm 0.05$ \\
NEOWISE/$W1$	  & 33,526    & $ 14.49    \pm 0.01$ \\
NEOWISE/$W2$	  & 46,028    & $ 15.13    \pm 0.03$ \\
\bottomrule    
\end{tabular}
\tablecomments{All measurements are reported in the AB system and are not corrected for reddening. For guidance, we report the effective wavelength of each filter.}
\end{table}